\documentclass{article}
\usepackage{graphicx} % Required for inserting images
\usepackage[margin=1in]{geometry}
\usepackage{graphicx}
\usepackage{amsmath}
\usepackage{array}
\usepackage{booktabs}
\usepackage{longtable}
\usepackage{multirow}
\usepackage{caption}
\usepackage{url}
\usepackage{hyperref}
\usepackage{algorithm2e} % Use this instead of the algorithmic packages
\usepackage{algorithmicx}
\usepackage{algpseudocode}
\usepackage{float}  
\usepackage{lipsum} % for dummy text

\providecommand{\keywords}[1]
{
  \small
  \textbf{\textit{Keywords—}} #1
}

\title{DCSNet: A Lightweight Knowledge Distillation-Based Model with Explainable AI for Lung Cancer Diagnosis from Histopathological Images}

\author{
  Sadman Sakib Alif\textsuperscript{1}, 
  Nasim Anzum Promise\textsuperscript{1}, 
  Fiaz Al Abid\textsuperscript{1}, 
  Aniqua Nusrat Zereen\textsuperscript{2,*} \\
  \textsuperscript{1}North South University, Dhaka, Bangladesh \\
  \textsuperscript{2}Mahidol University, Salaya, Thailand \\
  *Corresponding Author: Aniqua Nusrat Zereen, \texttt{aniqua.zer@mahidol.ac.th}
}

\begin{document}

\maketitle

\begin{abstract}
Lung cancer is a leading cause of cancer-related deaths globally, where early detection and accurate diagnosis are critical for improving survival rates. While deep learning, particularly convolutional neural networks (CNNs), has revolutionized medical image analysis by detecting subtle patterns indicative of early-stage lung cancer, its adoption faces challenges. These models are often computationally expensive and require significant resources, making them unsuitable for resource-constrained environments. Additionally, their lack of transparency hinders trust and broader adoption in sensitive fields like healthcare. Knowledge distillation addresses these challenges by transferring knowledge from large, complex models (teachers) to smaller, lightweight models (students). We propose a knowledge distillation-based approach for lung cancer detection, incorporating explainable AI (XAI) techniques to enhance model transparency. Eight CNNs, including ResNet50, EfficientNetB0, EfficientNetB3, and VGG16, are evaluated as teacher models. We developed and trained a lightweight student model, Distilled Custom Student Network (DCSNet) using ResNet50 as the teacher. This approach not only ensures high diagnostic performance in resource-constrained settings but also addresses transparency concerns, facilitating the adoption of AI-driven diagnostic tools in healthcare.
\end{abstract}


\section{Introduction}\label{sec1}

The lungs are one of the most vital organs in the human body, playing an essential role in the respiratory system. The lungs are also prone to several diseases, including lung cancer, which remains one of the deadliest forms of cancer worldwide. Lung cancer can be primarily divided into two groups: non-small cell lung cancer (NSCLC) and small-cell lung cancer (SCLC). About 80\% of lung cancer cases are of NSCLC. NSCLC is further categorized into subtypes such as adenocarcinoma (ACC), squamous cell carcinoma (SCC), adenosquamous carcinoma (AC) and sarcomatoid carcinoma (SC). ACC and SCC are two of the most common types of NSCLC. On the other hand, SCLC accounts for about 20\% of lung cancer cases. Small cell carcinoma and combined small cell carcinoma are two specific types of SCLC. Comparatively, SCLC grows more quickly and is harder to diagnose \cite{sutherland2010cell}.

As of 2024, lung cancer continues to be the most commonly diagnosed cancer and the leading cause of cancer related deaths, especially among men. It accounts for nearly 2.5 million cases, representing one in eight cancer diagnoses, and it causes approximately 1.8 million deaths \cite{a}. The risk of a man developing lung cancer in his lifetime is about one in 16 and for a woman it is one in 17.

Oncologists begin the process of diagnosing lung cancer by conducting physical examinations and blood tests to identify initial signs that may indicate the presence of lung cancer. Based on these results, advance diagnostic methods are carried out. These techniques can be categorized into non-invasive and invasive techniques. Non-invasive techniques include imaging methods such as X-rays, CT scans, MRIs and PET scans. A chest X-ray is used to detect the presence of masses or nodules, while CT scan provides detailed cross-sectional images, providing information about the size, shape and location of suspicious areas. 
MRI is effective in finding out the extent of cancer and its spread to other tissues. PET scans, on the other hand, detect cancer cells based on their increased metabolic activity. If these methods suggest the presence of cancerous cells, a lung biopsy is usually recommended. A lung biopsy is a procedure in which a sample of lung tissue is extracted for microscopic examination. There are several biopsy techniques such as needle biopsy, bronchoscopic biopsy, open biopsy and thoracoscopic biopsy. After performing the biopsy, the lung tissue is processed, stained and examined under a microscope to produce histopathological images. These histopathological images are crucial for pathologists to confirm and identify the specific type of lung cancer.

Artificial Intelligence (AI) based techniques are transforming the field of healthcare. Significant advancements in AI such as Deep Neural Networks (DNNs), convolutional neural networks (CNNs), Vision Transformers, and Generative Adversarial Networks (GANs) have enhanced the analysis of various medical images, such as PET scans, histopathological images, and MRIs. AI-based techniques have been extensively explored to be used for computer-aided detection (CAD) and classification of different types of diseases, such as tumors, strokes, cancers (including lung, breast, prostate, and colorectal cancers), heart diseases, neurological disorders (such as Alzheimer's disease, multiple sclerosis, and Parkinson's disease), cardiovascular diseases, fractures, etc. \cite{hosseini2024deep}, \cite{forte2022deep}. These AI-driven systems are capable of identifying complex underlying patterns and anomalies in medical images. CAD systems significantly minimize the possibility of human error, increasing the possibility of early detection of life-threatening diseases. These systems can process a vast amounts of imaging data quickly and in a reliable manner.

Model compression techniques, such as Knowledge Distillation (KD), have gained significant attention for their ability to reduce the complexity of deep learning models while maintaining high accuracy. KD is a form of model compression and acceleration where a smaller, less complex model, known as the student model, learns to approximate the behavior of a larger, more complex model, the teacher model. The overarching goal of KD is to enable the student model to generalize well while reducing computational and memory requirements, making it particularly useful in resource-constrained environments.

Knowledge Transfer in KD refers to the process by which the student model acquires knowledge from the teacher model. In this context, knowledge transfer is not limited to the explicit target labels used in traditional training but instead involves extracting useful information from the teacher model's outputs, intermediate features, or attention maps. This transfer of knowledge is typically performed using various strategies that dictate how the student model mimics the teacher model’s behavior.

The most widely adopted form of KD is response-based knowledge distillation, where the student model learns from the probability distribution of the teacher’s predictions rather than the ground truth labels. This approach enables the student to capture the subtle patterns learned by the teacher model, which might be overlooked in a traditional supervised learning setup. The objective is to minimize the divergence between the soft outputs (predicted probabilities) of the teacher and the student, often using a softmax temperature scaling technique to smooth the output distribution \cite{gou2021knowledge}.

Alternatively, feature-based knowledge distillation focuses on transferring knowledge from the intermediate feature representations within the hidden layers of the teacher model. The student is trained to replicate these internal representations, facilitating the student’s learning process by leveraging the teacher’s hierarchical feature extraction \cite{zhang2020improve}. This form of distillation aims to guide the student in learning more meaningful and structured features, thus improving its generalization performance.

Attention-based knowledge distillation further refines this process by aligning the attention maps of the teacher and student models. In this approach, the student model is encouraged to allocate attention to the same spatial regions or features that the teacher model deems important, which helps the student focus on the most relevant patterns in the input data \cite{ji2021show}. This can be particularly beneficial in tasks where certain spatial or feature-based information is crucial for accurate predictions, such as in medical image analysis.

Relation-based knowledge distillation takes a different approach by transferring the relational structure of the feature maps between the teacher and student models. In this case, the student model learns the relationships or dependencies between different features or spatial locations in the feature maps, which enables it to capture more intricate patterns and spatial relationships that are often learned by the teacher model \cite{park2019relational}.

By applying these various forms of knowledge distillation, the student model is able to effectively generalize while maintaining a compact architecture. This results in a compressed model that retains high accuracy, reduces memory usage, and lowers computational costs. Particularly in domains with limited computational resources, such as edge devices or embedded systems in medical image analysis, KD offers a viable solution for deploying AI models efficiently without sacrificing performance.

In our study, we propose a KD based approach for detecting lung cancer from histopathological images. Initially, we train a set of eight large CNNs, including ResNet50, ResNet101, EfficientNetB0, EfficientNetB3, VGG16, VGG19, MobileNetV2, and InceptionV3, using the LC25000 dataset \cite{borkowski2019lung}, consisting of a total of 15,000 histopathological images of cancerous and healthy lung tissue. Subsequently, we choose the best performing model as the teacher model. We then design a custom student model: Distilled Custom Student Network (DCSNet), which has less parameters. DCSNet learns from the output probabilities of the teacher. The student model is lightweight and it can capture nuanced class relationships, enhancing its generalization ability to unseen data. The DCSNet is able to identify different lung cancer cases, including ACC, SCC, and benign tissue (BN) from histopathological images. To enhance the interpretability of our model, we utilize Explainable AI (XAI) techniques. The main contributions of our paper are summarized as follows:\\

\begin{itemize}
    \setlength{\itemsep}{1em}

    \item We have provided a comprehensive empirical analysis to prove the reliability and the viability of the procedure in comparison within eight models: ResNet50, ResNet101, EfficientNetB0, EfficientNetB3, VGG16, VGG19, MobileNetV2, and InceptionV3, and identify the best teacher model.
    \item We have introduced a distilled compressed lightweight model: DCSNet specifically designed to perform well for histopathological images of lung cancer, contributing to the limited field of DL-based research on the disease.
    \item We have integrated XAI techniques to enhance the transparency and understanding of the model.\\

\end{itemize}

The remainder of the paper is organized as follows. In Section \ref{sec:related}, we provide an overview of related work on lung cancer detection using AI, studies that utilize the KD approach. Section \ref{sec:method} provides details about the proposed model. In Section \ref{sec:exp}, we provide a detailed analysis of the performance of our teacher and student models. Section \ref{sec:XAI} provides insights into the decision-making process of the proposed model (DCSNet) using XAI. Finally, Section \ref{sec:conclusion} summarizes the findings.

\section{RELATED WORKS}
\label{sec:related}

Researchers have explored various machine-learning approaches to diagnose lung cancer in patients. Previous studies have employed diverse features, including symptom-based diagnosis and medical imaging. In addition, many studies have also been done in the field of KD, where researchers have used many different variants of KD to build lightweight models. Several researchers have applied KD-based approaches to create efficient models for lung cancer detection. These studies show that applying KD can help to develop lightweight models while maintaining high accuracy.

\subsection{Lung Cancer Detection}
In recent years, several advances in the field of artificial intelligence have led to a rise in the analysis of medical images using machine learning techniques.

Nasser \textit{et al.} \cite{b1} proposed an Artificial Neural Network (ANN) system for the detection of lung cancer based on patient symptoms and personal information such as anxiety, chronic disease, fatigue, allergies, and other factors. The model achieved an accuracy level of 96.67 \%. Although the system demonstrates impressive performance, its predictions rely on accurate symptom data and personal information, which limits its generalization ability.
A different approach is proposed by Al-Tarawneh \cite{b2}. Instead of relying on patient-reported symptoms, CT scan images are analyzed using image processing techniques such as Gabor filtering, watershed segmentation, and pattern recognition. This method achieves an accuracy of 92.86\%, which is slightly lower compared to the previous method. But it is less dependent on patient-specific factors, making it robust. However, it has certain limitations, as the prediction accuracy depends on the quality of the images.
To address the issue of image quality, Shakeel \textit{et al.} \cite{b9} applied denoising to the input images to enhance the quality. The enhanced images were segmented using an improved profuse clustering technique, and the segments were then classified using deep neural networks.
Sori \textit{et al.} \cite{b10} also utilized denoising to enhance the quality of the CT scan images. They utilized a DFD-Net model, which employs a two-path CNN to integrate local and global features. Their detection accuracy is 96.66\% for correctly identifying lung cancer nodules. 
A multilevel brightness-preserving approach to eliminate noise and increase the quality of the CT scans is proposed by Shakeel \textit{et al.} \cite{b7}. The system utilized an improved deep neural network to segment the affected regions and extract various features, which are then classified using an ensemble classifier. This approach achieved an accuracy of 96\%.  
A similar approach is taken by Bhatia \textit{et al.} \cite{b6}, where the system identified regions of the lung most probable to have lung cancer before applying deep learning models such as ResNet and U-Net to extract features. These features were then analyzed using classifiers like XGBoost and Random Forest and achieved an accuracy of 84\% on the LIDC-IRDI dataset. The method proposed by Asuntha \textit{et al.} \cite{b4} employed multiple feature extraction and utilized fuzzy particle swarm optimization to select the optimal features. Both underscore the significance of feature narrowing to improve model performance.
Recent advancements have also incorporated 3D-CNN for detecting lung cancer from CT scans. In the method proposed by Alakwaa \textit{et al.} \cite{b8}, the lung tissue is segmented using thresholding and a modified U-Net is applied to identify potential nodules, followed by classification with a 3D-CNN. This approach achieved an accuracy of 86.6\%.

\subsection{Knowledge Distillation}
KD is widely used in recent studies, where smaller student models are trained to mimic the behavior of larger teacher models. Hinton \textit{et al.} \cite{hinton2015distilling}, showed the student learns to mimic the output logits of the teacher models.
Fukuda \textit{et al.} \cite{b11} proposed an ensemble KD approach, where a smaller student network is trained using a set of teacher models, such as VGG or LSTM. The student model simultaneously learns from the outputs of multiple teachers. 
Yoon \textit{et al.} \cite{b13} introduced a similarity-preserving knowledge distillation, where a small student network is trained to replicate the output of an extensive teacher network while maintaining important similarities between data.

These methods transfer knowledge from a large teacher to a smaller student model. However, if the student is significantly less complex it may fail to capture and replicate the teacher's intricate outputs.
This issue is addressed in the multi-step KD method. Mirzadeh \textit{et al.} \cite{b12} utilized the Teacher Assistant Knowledge Distillation (TAKD) approach. TAKD utilizes an intermediate model, called a Teacher Assistant (TA), to bridge the gap between a large teacher model and a small student model. Training a student using the TA gives better performance as the TA is closer in size to the student and makes the transfer of knowledge effective.
Saleknia \textit{et al.} \cite{saleknia2024multi} also employed mid-size assistant networks that bridge the computational gap between the teacher and student models. They use three variants of EfficientNetV2 to build the teacher, teacher assistant, and student networks. Their approach improved the student model performance on the Standford-40 dataset from 94.75\% to 96.30\%. A disadvantage of this method is that computational complexity increases because of the multi-step distillation process

However, transferring knowledge using only output logits has certain limitations, as it solely focuses on the output distribution while completely ignoring the intermediate representations learned at various layers. Feature-based distillation strategies address this limitation by transferring knowledge from intermediate layers. 
Remero \textit{et al.} \cite{romero2014fitnets} used both the output logits and the intermediate representations of the teacher model to guide the training process of the student. Additionally, they introduced extra parameters to map the student's hidden layers to the prediction of the teacher's hidden layers. Using this approach a student network with almost 10.4 times fewer parameters outperformed a larger teacher model.
Zagoruyko \textit{et al.} \cite{zagoruyko2016paying} utilized the attention maps of a powerful teacher network to train the student model. These attention maps represent the spatial maps within CNNs that encode the spatial areas the network focuses on when making decisions. They experimented using two types of attention maps: activation-based maps and gradient-based maps. Experimental results showed consistent improvement across various datasets and CNN architectures. Yang \textit{et al.} \cite{yang2024vitkd} proposed ViTKD, which mimics the shallow layers and generates the deep layer of the teacher model.
Guo \textit{et al.} \cite{b14} developed Class Attention Transfer Knowledge Distillation (CAT-KD), where student models are improved by transferring Class Activation Maps (CAMs) from teacher models. This method focuses the student model on key areas in input data to provide better classification.

So far, a common aspect of all of the methods discussed so far is they require a large teacher model to train the student model. This has some inherent limitations, such as the need for pre-training a complex teacher model. Also, the performance of the same student network can vary depending on the choice of the teacher network. 
To overcome this limitation, Zhang \textit{et al.} \cite{zhang2019your} proposed self-distillation, where the student network distills knowledge internally by transferring information from the deeper layers to the shallow ones. Their experiments show an average accuracy improvement of 2.65\%, with enhancements ranging from 0.6\% in ResNet to 4.07\% in VGG19. 
Xu \textit{et al.} \cite{xu2019data} introduced data-distortion guided self-distillation where knowledge is transferred within the same network by using different distorted versions of the same training data. This way the network learns consistent global features and class probabilities across different variations of the data. Experiments on CIFAR-10, CIFAR100, and ImageNet show this approach improves the performance of various architectures like AlexNet, ResNet, Wide ResNet and DenseNet. 

A lot of work has been done to enhance and improve the traditional framework of KD to enhance performance. 
For example, Tu \textit{et al.} \cite{tu2022general} proposed dynamic KD where teacher and student models learn from each other. Although this approach enhanced performance, it also increased the computational complexity due to continuous interaction between the models.
Yang \textit{et al.} \cite{yang2023skill} introduced the Skill-Transferring Knowledge Distillation (SKD) method, which consists of two meta-learning networks: Teacher Behavior Teaching and Teacher Experience Teaching. The first network captures how the teacher learns in its hidden layers and can predict its future behavior based on its past behavior. The latter network models the optimal knowledge acquired by the teacher network from its output layer at each stage of the learning process. With the help of meta-learning networks a teacher network can provide its actions to the student, which include the predictions of its future behavior and the optimal empirical knowledge from the current stage.
Zheng \textit{et al.} \cite{zhang2023adaptive} proposed adaptive multi-teacher Knowledge Distillation with meta-learning (MMKD). Their approach utilized a meta-weight network to distill knowledge from the output and intermediate layers of the teacher model. MMKD achieved a top-1 accuracy of 75.66\% on the CIFAR-100 dataset. Despite the superior performance, this approach increased computational complexity due to the meta-weight network.
Wang \textit{et al.} \cite{wang2022nffkd} fused multi-level features and applied hierarchical mixed loss (HML) to improve learning. Their approach achieved top-1 accuracy of 77.49\% on the CIFAR-100 dataset. 
Rao \textit{et al.} \cite{rao2023parameter} proposed Parameter-Efficient and student friendly Knowledge Distillation. This approach used adapter modules to update only a few parameters in the teacher model. The method was tested on various benchmarks and showed that PESF-KD could perform similarly or better than most other distillation methods.

\subsection{Knowledge Distillation in medical domain }
KD has tremendous potential, but it largely remains unexplored in the medical domain. However, some studies utilize the powerful model compression strategy.
Qin \textit{et al.} \cite{qin2021efficient} applied KD for medical image segmentation, transferring semantic region information from teacher to student network. By encoding internal information of each semantic region during the transfer process, their approach addressed the ambiguous boundary problem often encountered in medical imaging.
Zhao \textit{et al.} \cite{zhao2023mskd} proposed a structured distillation framework specifically designed for medical image segmentation. The framework consists of feature filtering distillation that transfers important knowledge from the teacher to the student while minimizing the influence of redundant information. Another component of the framework is region graph distillation which transfers structure-dependent information by leveraging advanced representational capabilities of graphs. Their results demonstrate improvements of up to 18.56\% in the Dice coefficient for lightweight neural networks.
Xing \textit{et al.} \cite{xing2021categorical} proposed class-guided contrastive distillation for medical image classification. Their approach moves positive image pairs closer in the feature space while pushing negative image pairs farther apart. This regularization strategy enhances intra-class similarity and increases inter-class variance. 
Hassan \textit{et al.} \cite{hassan2022knowledge} proposed knowledge-distillation based instance segmentation that allows conventional semantic segmentation models to identify stroma, benign and cancerous prostate tissue from whole slide images (WSIs) with incremental few-shot learning. They evaluate their method on two datasets, where their approach outperforms state of the art methods by 2.01\% and 4.45\% in mean IoU scores for prostate tissue segmentation, and 10.73\% and 11.42\% in F1-scores for grading prostate cancer (PCa)  as per clinical standards.

Only a few studies have explored the application of KD in lung cancer detection and classification.
Wang \textit{et al.} \cite{b25} showed how to improve the detection of lung nodules using elastic weight consolidation (EWC) and feature distillation. This method helps a model learn new data while remembering past knowledge, reducing the need for storing all past data. Their method achieved a 2\% improvement in accuracy on the LUNA16 dataset.
Zheng \textit{et al.} \cite{b26} proposed KD\_ConvNeXt, where they classified lung tumor subtypes using KD. The model uses a ConvNeXt student network to learn from a Swin Transformer teacher network, which helps the student network focus on important features in the data. Their method achieved a classification accuracy of 85.64\%, and F1-score of 77.17\%. 
Table~\ref{tab:related_summary} shows the summary of related works on lung cancer detection and KD.

\pagebreak

\begin{table*}[ht]
\centering
\caption{\centering Summary of the related works}
\label{tab:related_summary}
\renewcommand{\arraystretch}{1.5}
\resizebox{\textwidth}{!}{%
\begin{tabular}{@{}p{2cm}p{6cm}p{4cm}p{5cm}@{}}
\toprule
\textbf{Authors} & \textbf{Methods and Features} & \textbf{Achievements} & \textbf{Limitations} \\
\midrule
Nasser \textit{et al.} \cite{b1} & ANN system for lung cancer detection using patient symptoms and personal information & Accuracy: 96.67\% & Predictions rely on accurate symptom data and personal information \\
\midrule
Al-Tarawneh \cite{b2} & CT scan image analysis using Gabor filtering, watershed segmentation, and pattern recognition & Accuracy: 92.86\% & Dependent on image quality \\
\midrule
Sori \textit{et al.} \cite{b10} & Denoising, DFD-Net model with two-path CNN to integrate local and global features & Accuracy: 96.66\% for identifying lung cancer nodules & Loss of image details due to denoising with DR-Net, which affects accuracy \\
\midrule
Bhatia \textit{et al.} \cite{b6} & ResNet and U-Net for feature extraction, XGBoost and Random Forest for classification & Accuracy: 84\% on LIDC-IRDI dataset & Lower accuracy compared to other methods \\
\midrule
Alakwaa \textit{et al.} \cite{b8} & 3D-CNN, thresholding, and modified U-Net for lung tissue segmentation and nodule identification & Accuracy: 86.6\% & Can detect cancer but cannot determine the exact location of the cancerous nodules \\
\midrule
Saleknia \textit{et al.} \cite{saleknia2024multi} & Multi-step KD using EfficientNetV2 variants for teacher, assistant, and student networks & Improved accuracy on Stanford-40 dataset from 94.75\% to 96.30\% & Performed well but increased computational complexity \\
\midrule
Romero \textit{et al.} \cite{romero2014fitnets} & Feature-based KD using intermediate representations and additional parameters & Student model with 10.4 times fewer parameters outperformed teacher & Suggested more efforts should be devoted to exploring new training strategies \\
\midrule
Zheng \textit{et al.} \cite{zhang2023adaptive} & Adaptive multi-teacher KD with meta-learning (MMKD) using meta-weight networks & Accuracy: 75.66\% on CIFAR-100 & Lower accuracy \\
\midrule
Wang and Luo \cite{b25} & Elastic Weight Consolidation (EWC) and Feature Distillation to improve lung nodule detection by learning new data while retaining past knowledge & Increased accuracy on LUNA16 dataset by 2\% over previous methods & Limited to 2D lung nodule detection and EWC algorithm bounding box regression learning is poor \\
\midrule
Zheng \textit{et al.} \cite{b26} & KD\_ConvNeXt: KD where a ConvNeXt student network learns from a Swin Transformer teacher network for lung tumor subtype classification & Classification accuracy: 85.64\%, F1-score: 0.7717 & Effectiveness of the model still needs improvement, and classification accuracy is poor \\
\bottomrule
\end{tabular}%
}
\end{table*}

\section{Methodology}
\label{sec:method}
This section outlines the details of the proposed approach from setup to evaluation. The system setup describes the computational resources and configurations used for the experiments. Next, we discuss the dataset and its characteristics. This is followed by a detailed explanation of the KD technique implemented in our work. We then describe the model training, testing and performance evaluation.

\subsection{System Setup}
\begin{figure}[ht]
    \centering
    \includegraphics[width=\textwidth]{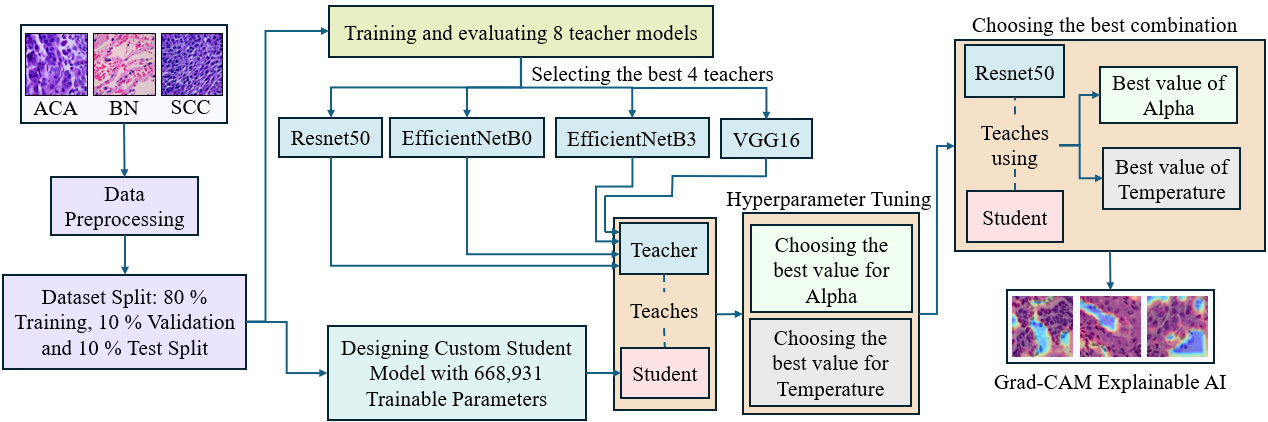} % Use full text width
    \caption{\centering Workflow of the proposed system}
    \label{fig:workflow}
\end{figure}

All of our experiments are carried out on the Kaggle platform, leveraging its computational resources for training and evaluation. For training, we utilize Kaggle's Nvidia Tesla P100 GPU and 16 GB of RAM. Fig. \ref{fig:workflow} illustrates the entire workflow of our proposed methodology, highlighting each step of our approach.
\subsection{Dataset details}
\begin{figure}[ht]
    \centering
    \includegraphics[width=0.9\textwidth]{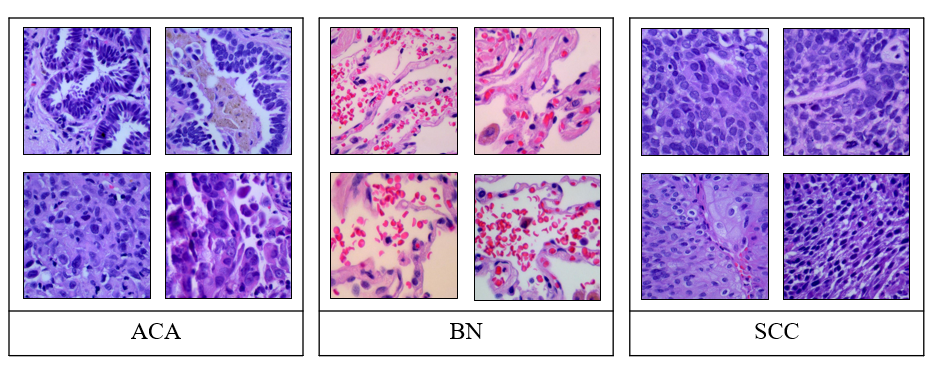}
    \caption{\centering Images of three types of lung cancer}
    \label{fig:dataset_samples}
\end{figure}

The dataset used in our work is collected from the study conducted by Andrew A. Borkowski, MD and his team \cite{borkowski2019lung}. 
Originally the researchers collected 750 color histopathological images of lung tissue (250 BN, 250 ACC, and 250 SCCs) using a Leica Microscope MC190 HD camera connected to an Olympus BX41 microscope and the Leica Acquire 9072 software for Apple computers. The original images are captured at a resolution of $1024 \times 768$ pixels from pathology glass slides using a 2012 Apple Macbook Pro. All the images are resized to $768\times768$ pixels. To expand the dataset, the researchers used the Augmentor software package, which is an image augmentation library in python used for machine learning tasks. The library is platform and framework independent and employs a stochastic approach using modular building blocks that can be connected together to form a sequential pipeline of operations \cite{bloice2019biomedical}. 
The total number of lung cancer images after augmentation is 15,000, divided evenly among the three classes: ACA (5,000), SCC (5,000), and BN (5,000). The augmentation strategies are, left and right rotations (up to 25 degrees with a probability of 1.0) and horizontal and vertical flips (with a probability of 0.5). 
Our training set contains 12,000 images (80\%of the total dataset), with 4,000 images from each class. The validation and test set each contains 500 images per class, summing up to 1,500 images (10\% of the total dataset) for each subset. Fig. \ref{fig:dataset_samples} shows sample images of each class from the dataset.

\newpage
%topics:KD, types of KD, our implemented kd type
\subsection{Knowledge Distillation}
\begin{figure}[ht]
    \centering
    \includegraphics[width=\textwidth]{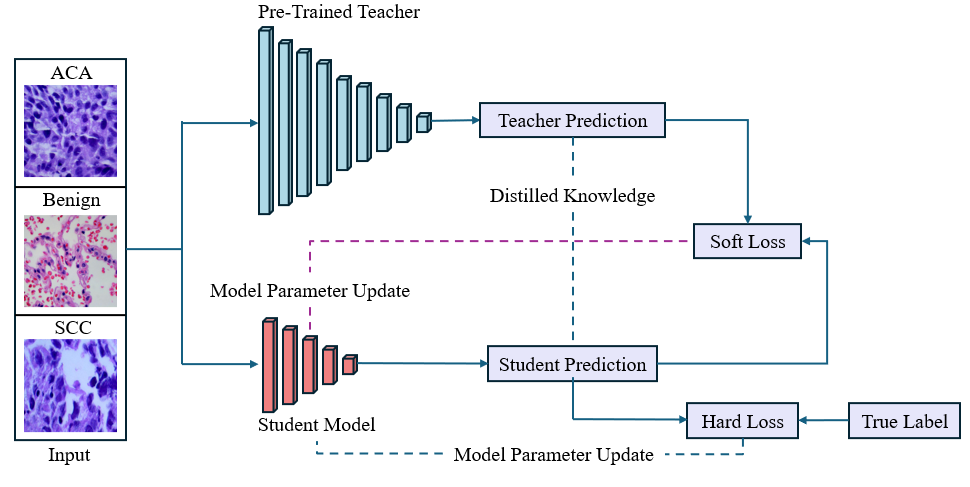}
    \caption{\centering Student model learning from teacher model and ground truth (True Label)}
    \label{fig:kd framework}
\end{figure}

In this work, response-based KD is utilized. Fig. \ref{fig:kd framework} shows the framework of the KD approach. In response-based KD, the teacher's knowledge is directly transferred using the teacher model's output predictions, essentially teaching the student model to mimic the teacher's response from the probability distribution over the output classes. The main idea is to use the softened outputs of the teacher. These softened outputs are calculated by first dividing the output logits (raw outputs prior to applying the softmax function) by a temperature scaling factor and then applying the softmax function. This process helps to control the smoothness of the output probability distribution. For a class $i$, the softened outputs are calculated using equation \ref{eq:softened_outputs},

\begin{equation}
P_i = \frac{\exp\left(\frac{z_i}{T}\right)}{\sum_{j} \exp\left(\frac{z_j}{T}\right)}
\label{eq:softened_outputs}
\end{equation}

Here, $P_i$ represents the softened probability for class i, $z_i$ is the raw output for class $i$ before applying the softmax function, and $j$ is an index that iterates over all the classes. The temperature parameter $T$ controls the output's smoothness. A higher value of $T$, results in a smoother distribution, while a lower value of $T$ results in a much more sharp distribution. The softened probability scores represent a richer distribution over the classes, capturing the probable class and also the relative probabilities of less likely classes. The total distillation loss in KD integrates two components: the soft loss and the hard loss. The soft loss measures how well the student model approximates the softened probability distributions of the teacher. It is calculated using equation \ref{eq:loss_soft}.

\begin{equation}
\mathcal{L}_{\text{soft}} = -\sum_{i} p_{\text{teacher}}(x_i) \log(p_{i, \text{student}})
\label{eq:loss_soft}
\end{equation}

Here, \(p_{i, \text{student}}\) and \(p_{\text{teacher}}\) are the softened outputs of the teacher and student models, respectively. The hard loss measures how well the student model's predictions match the true labels. It is calculated using equation \ref{eq:loss_hard}, which is also known as the categorical cross-entropy loss.

\begin{equation}
\mathcal{L}_{\text{hard}} = -\sum_{i} y_i \log(p_{\text{student}}(x_i))
\label{eq:loss_hard}
\end{equation}

Here, $y_{i}$ represents the true label for the i-th sample, $log(p_{student}(x_{i}))$ is the natural logarithm of the predicted probability for the correct class.
The total loss is calculated as a weighted combination of these two components. It is calculated using equation \ref{total_loss_kd} :

\begin{equation}
\mathcal{L}_{\text{total}} = \alpha \mathcal{L}_{\text{hard}} + (1 - \alpha) \mathcal{L}_{\text{soft}}
\label{total_loss_kd}
\end{equation}

Here, \(\mathcal{L}_{\text{hard}}\) represents the loss based on the true labels, quantifying how well the student model predicts the actual data. Conversely, \(\mathcal{L}_{\text{soft}}\) captures the knowledge distilled from the teacher's output distributions. The hyperparameter \(\alpha\) balances the importance of these two losses. By fine-tuning this parameter, we control the focus of the training process.

\subsubsection{Teacher Models}
In the proposed approach, eight different traditional architectures are used which include: ResNet50, ResNet101, EfficientNetB0, EfficientNetB3, VGG16, VGG19, MobileNetV2, and InceptionV3, as potential teacher models. Further details on these architectures are provided in the following section.
\begin{itemize}
    \item \textbf{ResNet50}:
         ResNet50 is a part of the Residual Network architecture, which consists of five convolutional layers, including 16 residual blocks. It has a total of 50 deep layers containing 48 convolution layers, one Max Pooling layer, and one Average Pooling layer. Our ResNet50 model has almost 50 million parameters. ResNet is well known for its ability to reduce the vanishing gradient problem.  It also provides the best accuracy possible in some of the existing problems \cite{DBLP:journals/corr/HeZRS15}.
    \item \textbf{ResNet101}:
        ResNet101 is an extension of ResNet50 with some additional layers. It is a 101-layer residual network. The total number of parameters for our ResNet101 model is almost 68 million. As ResNet101 has many parameters, it is computationally expensive and more powerful \cite{DBLP:journals/corr/HeZRS15}.
    \item \textbf{EfficientNetB0}:
        EfficientNetB0 contains approximately 20M parameters. It is constructed using 16 inverted Residual Blocks split into seven stages. It is a computationally efficient model that can outperform larger models like ResNet, Inception, etc. EfficientNetB0 uses the Swish activation Function known as Sigmoid Linear Unit (SiLU), Global Average Pooling, and Softmax activation function. The SiLU (Sigmoid Linear Unit) function and Softmax activation function are defined as:
        \begin{equation}
        \label{eq:silu}
        \text{SiLU}(x) = x \cdot \sigma(x)
        \end{equation}
        \begin{equation}
        \label{eq:sig}
        \sigma(x) = \frac{1}{1 + e^{-x}}
        \end{equation}

        In equation \ref{eq:silu}, $x$ is the input value and $\sigma(x)$ is the sigmoid function. Equation \ref{eq:sig} shows the sigmoid function, which maps the input $x$ to a value between 0 and 1 \cite{DBLP:journals/corr/abs-1905-11946}.

    \item \textbf{EfficientNetB3}:
        EfficientNetB3 has approximately 30M parameters. It is an extension of EfficientNetB0 with more layers and higher input resolution. This model uses more resources, including compound scaling, to achieve better performance. Compound scaling helps the model to balance between accuracy and efficiency \cite{DBLP:journals/corr/abs-1905-11946}.
    \item \textbf{VGG16}:
        The VGG16 consists of 13 convolution layers and three fully connected layers, 16 layers in total with five pooling layers. There are five blocks in this architecture. Each of them contains two or three convolution layers and a max pooling layer of size 2 $\times$ 2. Following the last Max Pooling layer, there are three fully connected layers and a Softmax activation function to predict the output \cite{simonyan2014very}.
    \item \textbf{VGG19}:
        VGG19 is an extended version of VGG16 consisting of 16 convolutional layers and three fully connected layers.  Like VGG16, it also has five blocks and five pooling layers. But instead of two or three, there are two or four convolution layers in each of the blocks. VGG19 has more parameters and a deeper convolutional neural network than VGG16 \cite{simonyan2014very}.
    \item \textbf{MobileNetV2}:
        MobileNetV2 is a lightweight model developed for mobile and embedded vision applications. ModileNetV2 expands MobileNetV1 with multiple inverted residual blocks and linear bottleneck modules.  It is an efficient model for getting good performance regarding computational cost and memory usage \cite{DBLP:journals/corr/HowardZCKWWAA17}.
    \item \textbf{InceptionV3}:
        InceptionV3 is a CNN model that consists of 48 layers. It uses batch normalization extensively, along with multiple activation functions. A convolution kernel splitting method, Factorized Convolution, is one of the core parts of InceptionV3, which helps to reduce computational cost. Factorized Convolution splits a 3 $\times$ 3 convolution into 3$\times$1 and 1$\times$3 convolution. InceptionV3 architecture has almost 35M parameters \cite{DBLP:journals/corr/SzegedyLJSRAEVR14}.
\end{itemize}

\subsubsection{Student Models}
\begin{figure}[ht]
    \centering
    \includegraphics[width=\textwidth]{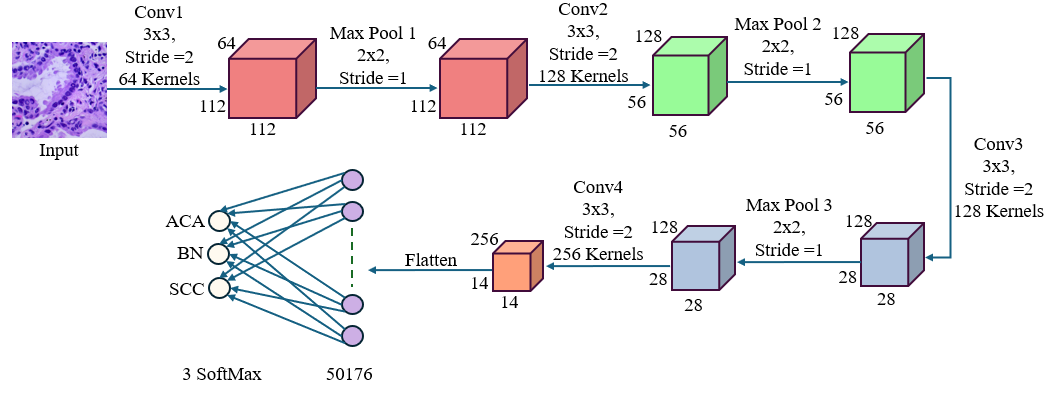}
    \caption{\centering DCSNet Architecture}
    \label{fig:sm_arch}
\end{figure}

We design our own custom model: DCSNet. In the following section, we delve into the detailed architectural description of DCSNet:

\textbf{DCSNet:} The model's input layer takes images of shape (\(224 \times 224 \times 3\)). The first convolutional layer consists of 64 filters of size (\( 3 \times 3 \)) with a stride of (\( 2 \times 2 \)), followed by a LeakyReLU activation function with an alpha value of 0.2. Next, a MaxPooling2D layer with a pool size of (\( 2 \times 2 \)) and strides of (\( 1 \times 1 \)) is utilized, which downsamples the feature maps while preserving the spatial dimensions as much as possible. This pattern is repeated for the next two convolution layers, with the number of filters increasing to 128. The final convolution layer consists of 256 filters. This gradual increase in the number of filters allows the model to capture more complex features in deeper layers. MaxPooling2D layer is followed by a dropout layer with a rate of 0.25 to reduce the risk of overfitting by randomly setting the fraction of the input units to zero during training. 
Finally, a flatten layer transforms the 2D feature maps into a 1D vector, which is subsequently passed into a dense layer with three output neurons and a softmax activation function.
Fig. \ref{fig:sm_arch} shows a detailed description of DCSNet.

\subsection{Model Training}
We train our models using the Adam optimizer with a learning rate of 0.001 and a batch size of 128. The teacher network's loss is computed using the categorical cross-entropy loss function. The distillation loss is calculated using Kullback-Leibler (KL) divergence, and the student network's loss is calculated using categorical cross-entropy.

\subsubsection{Algorithm}
Algorithm \ref{kd_algo} shows the pseudocode of the KD process. The input images are resized to a standard shape of ($224\times224\times3$). Then, a pre-trained model is selected and trained on the dataset as the teacher model. The output logits of the trained teacher are extracted, and $T$ is applied to generate softened probability distributions. A balancing factor $\alpha$ is set to control the contribution of the teacher’s knowledge in the distillation process. Next, a student model is built, and a distillation loss function is defined. The student model is then iteratively trained using the distillation loss function.

\begin{algorithm}[h]

\SetAlgoLined
\KwIn{$\mathcal{D} = \{(x_1, y_1), \ldots, (x_n, y_n)\}$: $\mathcal{D}$ represents the training set, where $x_i$ denotes the input and $y_i$ is the class label of $x_i$}
\KwOut{$\hat{y}$ contains the predicted class labels.}

\For{$i = 1$ to $n$} {
    Assemble each histopathological image in terms of classification label\;
    Resize images by $224 \times 224 \times 3$\;
}
Import pre-trained teacher model\;
Train teacher model on data\;

\For{$i = 1$ to $n$} {
    Training teacher\;
}

\textbf{Distill knowledge:} \\
Extract logits from teacher model\;
Select temperature $T$ and alpha $\alpha$\;
Apply softmax with $T$ and get soften probabilities\;
Build student model\;
Define distillation loss\;

\For{$i = 1$ to $n$} {
    Training student\;
}

\caption{Pseudocode of Knowledge Distillation}
\label{kd_algo}
\end{algorithm}

\subsubsection{Hyperparameter Tuning}
Hyperparameters such as $T$ and \( \alpha \) contribute directly to the quality of the distillation process. In KD, the teacher model makes predictions in the form of logits (raw outputs prior to applying the softmax function. The softmax function converts these logits into probability distributions, representing the likelihood of each class. As shown in equation \ref{eq:softened_outputs}, the $T$ modifies the softmax function by dividing the logits by the temperature value before applying the exponential function. 
Initially, we perform a random search for the value of $T$ within the range of 1 to 100, using large intervals between values. For example, we experimented with values such as 10, 20, 40, 60 and 90. This approach allowed us to cover a wide range of temperature settings and identify regions where the performance peaked. After narrowing down the search space, we conducted a more concentrated search around that region to fine-tune the temperature selection. 

By varying the value of $\alpha$, the importance of the distillation loss relative to the classification loss can be adjusted. In the experiment, the value of $\alpha$ is varied between 0.1 to 0.5 along with different temperature settings. This variation in $\alpha$ is applied thoroughly for all candidate pairs of teacher-student.

\subsection{Performance Metrics}
We utilize several performance metrics to evaluate the effectiveness of the proposed approach.
These metrics include Accuracy, Precision, Recall, and F1-score utilizing the following equations for True Positives (TP), True Negatives (TN), False Positives (FP), and False Negatives (FN). Accuracy is calculated using equation \ref{eq:accuracy}.

\begin{equation}
\label{eq:accuracy}
\text{Accuracy} = \frac{\text{TP} + \text{TN}}{\text{TP} + \text{TN} + \text{FP} + \text{FN}}
\end{equation}

Precision indicates the proportion of true positive predictions out of all instances that are predicted as positive, as shown in equation \ref{eq:precision}.

\begin{equation}
\label{eq:precision}
\text{Precision} = \frac{\text{TP}}{\text{TP} + \text{FP}}
\end{equation}

Recall, also called sensitivity, represents the percentage of true positive cases accurately detected by the model. It is essential in situations where reducing false negatives is crucial
Equation \ref{eq:recall} shows the formula for calculating recall.

\begin{equation}
\label{eq:recall}
\text{Recall} = \frac{\text{TP}}{\text{TP} + \text{FN}}
\end{equation}

The harmonic mean of precision and recall is denoted as F1-score. Equation \ref{eq:f1_form} demonstrates the formula for calculating F1-score.

\begin{equation}
\label{eq:f1_form}
\text{F1-score} = 2 \times \frac{\text{Precision} \times \text{Recall}}{\text{Precision} + \text{Recall}}
\end{equation}

\subsection{Proposed Teacher-Student Combination}
Through experimentation, we find that the optimal teacher for DCSNet is ResNet50. This selection is based on multiple evaluation metrics, such as accuracy, precision, recall, F1-score, and the number of parameters.
We choose ResNet50 as our teacher network due to its superior performance in understanding intricate patterns within the input data and effectively distilling that knowledge to the student network, outperforming other evaluated teacher models. For the student, our designed DCSNet offers a significant reduction in parameter count without substantial compromise in performance. Additionally, we also considered some pre-trained models as student models for comparison and observed that, although DCSNet sometimes has slightly lower accuracy than some of these models, the gain in efficiency outweighs the performance trade-off. 

\section{Experiments and Result analysis}
\label{sec:exp}
In this section, the performance of the candidate teacher models, assessing their effectiveness based on various metrics such as accuracy, precision, recall, and F1-score are described. Next, the training of the proposed student model using the best-performing teacher models are shown. This analysis includes a detailed examination of the effects of $T$ and \(\alpha\) on the distillation process.

\subsection{Model testing}

\begin{figure}[H]
    \centering

    \begin{minipage}{0.3\linewidth}
        \centering
        \includegraphics[width=\linewidth]{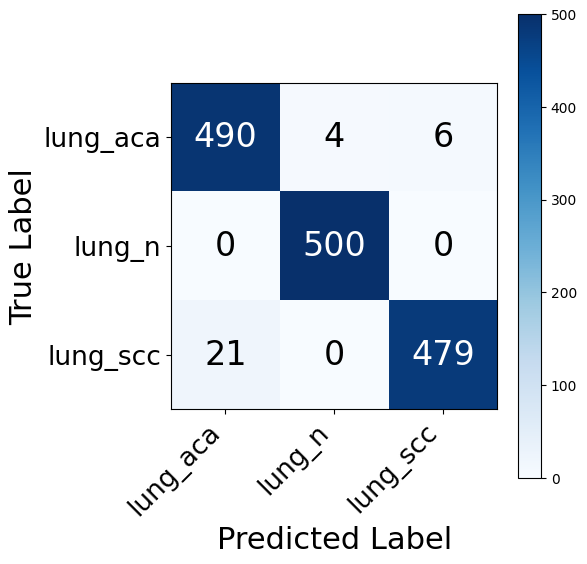}
        \\ (a) ResNet50
        \label{fig:Resnet50}
    \end{minipage}
    \begin{minipage}{0.3\linewidth}
        \centering
        \includegraphics[width=\linewidth]{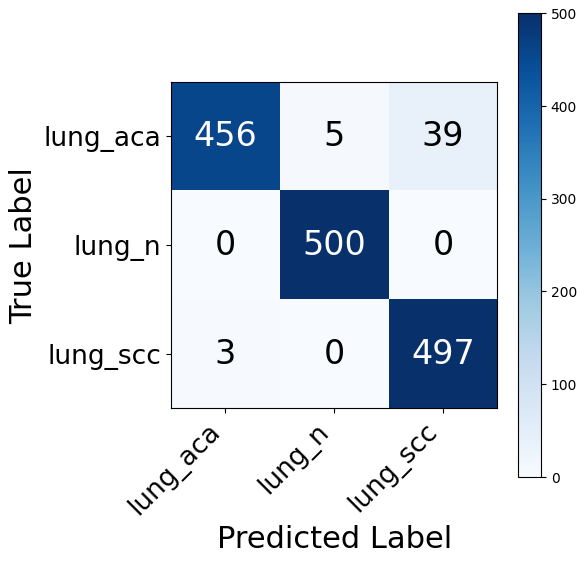}
        \\ (b) ResNet101
        \label{fig:Resnet101}
    \end{minipage}
    \begin{minipage}{0.3\linewidth}
        \centering
        \includegraphics[width=\linewidth]{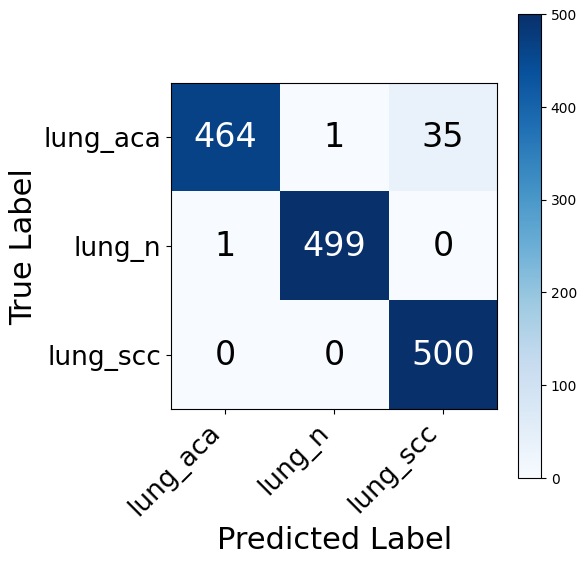}
        \\ (c) EfficientNetB0
        \label{fig:EfficientNetB0}
    \end{minipage}

    \vspace{0.5cm} % Add vertical space between rows

    \begin{minipage}{0.3\linewidth}
        \centering
        \includegraphics[width=\linewidth]{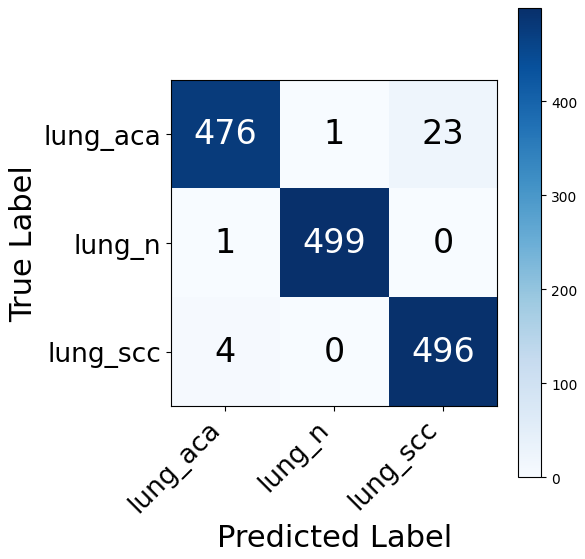}
        \\ (d) EfficientNetB3
        \label{fig:EfficientNetB3}
    \end{minipage}
    \begin{minipage}{0.3\linewidth}
        \centering
        \includegraphics[width=\linewidth]{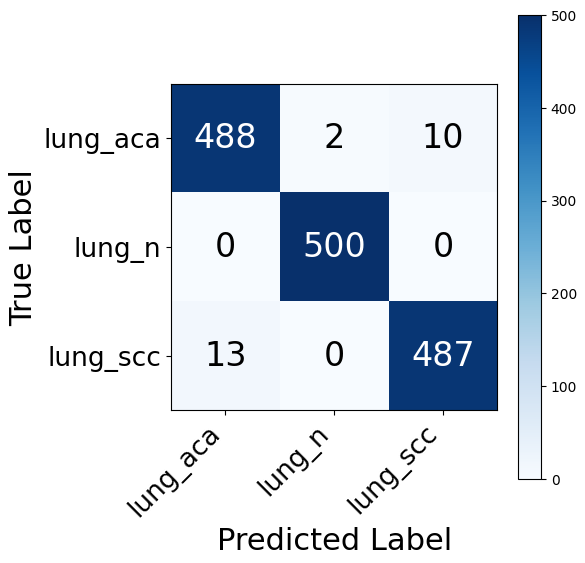}
        \\ (e) VGG16
        \label{fig:VGG16}
    \end{minipage}
    \begin{minipage}{0.3\linewidth}
        \centering
        \includegraphics[width=\linewidth]{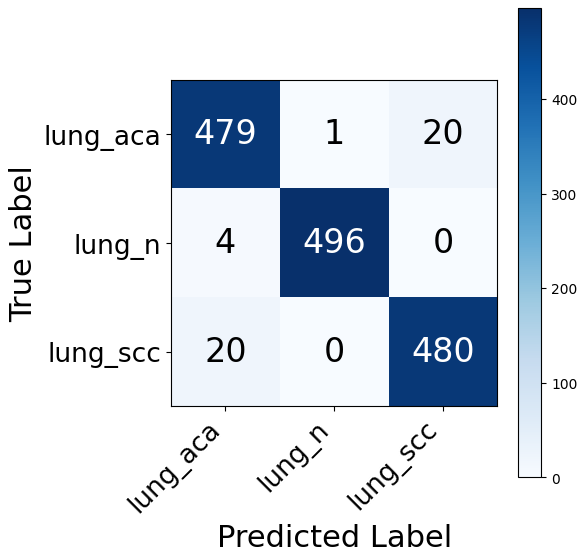}
        \\ (f) VGG19
        \label{fig:VGG19}
    \end{minipage}

    \vspace{0.5cm} % Add vertical space between rows

    \begin{minipage}{0.3\linewidth}
        \centering
        \includegraphics[width=\linewidth]{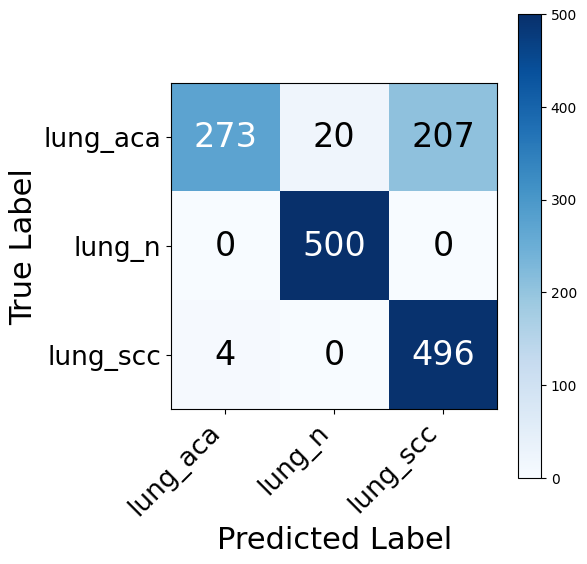}
        \\ (g) MobileNetV2
        \label{fig:MobilenetV2}
    \end{minipage}
    \begin{minipage}{0.3\linewidth}
        \centering
        \includegraphics[width=\linewidth]{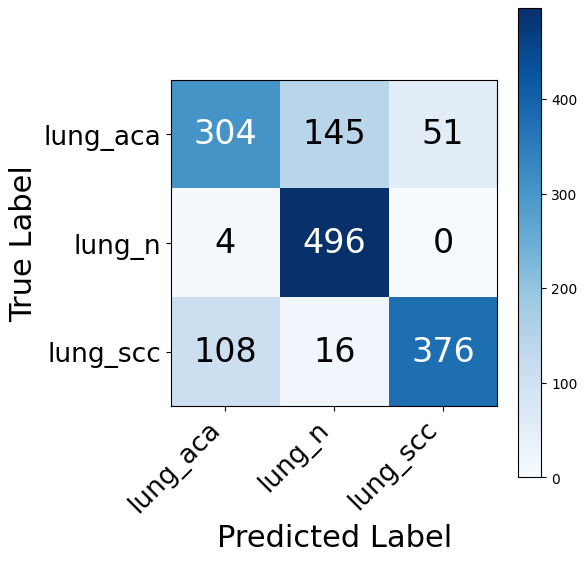}
        \\ (h) InceptionV3
        \label{fig:InceptionV3}
    \end{minipage}
\newline
    \caption{\centering Confusion Matrices of eight teacher models}
    \label{fig:CM_8Teachers}
\end{figure}

\textbf{Teacher performance.} Table \ref{tab:teacher_model_eval1} shows the performance of the different candidate teacher models. The results of the table highlight the differences in accuracy and complexity among different candidate teacher architectures. Models such as ResNet50 (49.31M), EfficientNetB0 (20.14M), EfficientNetB3 (30.08M), and VGG16 (21.17M) demonstrate the highest performance, each attaining accuracy, precision, recall, and F1-scores of 0.98. Conversely, MobileNetV2 (18.35M) and InceptionV3 (34.94M) achieve low accuracy scores of 0.85 and 0.78, respectively. These results demonstrate that although lightweight models such as MobileNetV2 is a good choice in terms of efficacy, they may suffer from performance issues due to their low parameter count. Additionally, the models with higher parameters, such as ResNet101 (68.38M) and VGG19 (26.48M), fail to significantly outperform their counterparts with fewer parameters, such as ResNet50 (49.31M) and VGG16 (21.17M). This suggests, an increase in model complexity does not necessarily correlate with improved performance. For further analysis of the performance of the teacher models, the confusion matrices of each candidate teacher are shown in Fig. \ref{fig:CM_8Teachers}. 

\begin{table}[ht]
\centering
\caption{\centering Teacher Model Evaluation}
\renewcommand{\arraystretch}{1.5}
\resizebox{\textwidth}{!}{%
\begin{tabular}{@{}lccccc@{}}
\toprule
\textbf{Model}          & \textbf{Parameter Number} & \textbf{Accuracy} & \textbf{Precision} & \textbf{Recall} & \textbf{F1-score} \\
\midrule
Resnet50        & 49,311,363 & 0.98 & 0.98 & 0.98 & 0.98 \\
Resnet101       & 68,381,827 & 0.97 & 0.97 & 0.97 & 0.97 \\
EfficientNetB0  & 20,139,430 & 0.98 & 0.98 & 0.98 & 0.98 \\
EfficientNetB3  & 30,084,658 & 0.98 & 0.98 & 0.98 & 0.98 \\
VGG16           & 21,170,755 & 0.98 & 0.98 & 0.98 & 0.98 \\
VGG19           & 26,480,451 & 0.97 & 0.97 & 0.97 & 0.97 \\
MobileNetV2    & 18,347,843 & 0.85 & 0.88 & 0.85 & 0.84 \\
InceptionV3    & 34,943,523 & 0.78 & 0.78 & 0.78 & 0.78 \\
\bottomrule
\end{tabular}%
}
\label{tab:teacher_model_eval1}
\end{table}

\begin{table}[ht]
\centering
\renewcommand{\arraystretch}{1.5}
\caption{\centering DCSNet Evaluation with ResNet50 as Teacher}
\resizebox{\textwidth}{!}{%
\begin{tabular}{@{}c c c c c c c c@{}}
\toprule
\textbf{Student Model} & \textbf{Trainable Parameter} & \textbf{Alpha} & \textbf{Temperature} & \textbf{Accuracy} & \textbf{Precision} & \textbf{Recall} & \textbf{F1-score} \\
\midrule

\multirow{10}{*}{DCSNet} 
& \multirow{10}{*}{668,931} 
& \multirow{10}{*}{0.3} 
& 7  & 0.89 & 0.90 & 0.89 & 0.89 \\ 
& &  & 10 & 0.92 & 0.92 & 0.92 & 0.92 \\ 
& &  & 20 & 0.86 & 0.87 & 0.86 & 0.86 \\
& &  & 40 & 0.92 & 0.92 & 0.92 & 0.92 \\
& &  & 46 & 0.87 & 0.88 & 0.87 & 0.86 \\ 
& &  & 47 & 0.91 & 0.91 & 0.91 & 0.91 \\
& &  & 50 & 0.89 & 0.89 & 0.89 & 0.89 \\ 
& &  & 61 & 0.87 & 0.90 & 0.87 & 0.87 \\ 
& &  & 62 & 0.89 & 0.89 & 0.89 & 0.89 \\ 
& &  & 64 & 0.89 & 0.89 & 0.89 & 0.89 \\ 

\bottomrule
\end{tabular}%
}
\label{tab:resnet_std}
\end{table}

\begin{table*}[ht]
\centering 
\caption{ \centering DCSNet Evaluation with VGG16 as Teacher}
\renewcommand{\arraystretch}{1.5}
\resizebox{\textwidth}{!}{%
\begin{tabular}{@{}c c c c c c c c@{}}
\toprule
\textbf{Student Model} & \textbf{Trainable Parameter} & \textbf{Alpha} & \textbf{Temperature} & \textbf{Accuracy} & \textbf{Precision} & \textbf{Recall} & \textbf{F1-score} \\
\midrule

\multirow{10}{*}{DCSNet} 
& \multirow{10}{*}{668,931} 
& \multirow{10}{*}{0.3} 
& 7   & 0.91 & 0.91 & 0.91 & 0.91 \\ 
&  &  & 12  & 0.75 & 0.83 & 0.75 & 0.71 \\ 
&  &  & 48  & 0.91 & 0.91 & 0.91 & 0.91 \\
&  &  & 53  & 0.89 & 0.90 & 0.89 & 0.89 \\
&  &  & 59  & 0.71 & 0.82 & 0.71 & 0.64 \\
&  &  & 72  & 0.66 & 0.83 & 0.66 & 0.55 \\ 
&  &  & 93  & 0.85 & 0.86 & 0.85 & 0.85 \\
&  &  & 96  & 0.89 & 0.89 & 0.89 & 0.89 \\ 
&  &  & 98  & 0.70 & 0.74 & 0.70 & 0.69 \\ 
&  &  & 100 & 0.81 & 0.85 & 0.81 & 0.80 \\ 

\bottomrule
\end{tabular}%
}
\label{tab:vgg_student}
\end{table*}

In the context of medical diseases, one factor is particularly critical: the FN rate in cancer detection. FNs occur when a test incorrectly identifies a cancerous condition as negative, leading to missed opportunities for early intervention. All top-performing models from Table \ref{tab:teacher_model_eval1}, including ResNet50, EfficientNetB0, EfficientNetB3, and VGG16, shows low FNs. ResNet50 has only four FNs for ACA and zero FNs for SCC. EfficientNetB0 and EfficientNetB3 each has one FN for ACA and zero FNs for SCC, while VGG16 has two FNs for ACC and zero FNs for SCC. In contrast, low-performing models such as MobileNetV2 and InceptionV3 show relatively high FN rates. InceptionV3 shows the highest rates of 145 FNs for ACA and 16 FNs for SCC.

From the analysis, the search for the optimal teacher models narrows down to four potential architectures: ResNet50, EfficientNetB0, EfficientNetB3, and VGG16. In the subsequent phase, each of these models are evaluated as a teacher for the student model. 

\textbf{Student Performance.}
In this section, we evaluate the performance of the proposed DCSNet. The evaluation is conducted using the test dataset, focusing on comparing student training with different teacher models and various configurations of $T$ and $\alpha$. Tables \ref{tab:resnet_std}, \ref{tab:vgg_student}, \ref{tab:effB0_stud}, and \ref{tab:effB3_stud} show the performance of the student model for different $T$ and $\alpha$ values when trained using the teacher models: ResNet50, VGG16, EfficientNetB0, and EfficientNetB3, respectively.

Table \ref{tab:resnet_std} illustrates the performance of the student model under varying $T$ values, using ResNet50 as the teacher model. The hyperparameter $\alpha$ is set to 0.3. From the data, we can see that varying the distillation temperature can have a significant effect on model performance. Notably, $T$ values of 10 and 40 achieve the highest accuracy, precision, recall, and F1-score, of 0.92. The lowest accuracy is 0.86, recorded at a $T$ of 20. Table \ref{tab:vgg_student} shows the performance of our student model for different $T$ values, using VGG16 as the teacher model and an $\alpha$ value of 0.3. The student model attains a peak accuracy of 0.91 at $T$ values of 7 and 48. Conversely, the lowest accuracy is 0.66, at a $T$ of 72. Table \ref{tab:effB0_stud} shows the performance of the student model with EfficientNetB0 as the teacher with $\alpha$ set to 0.4. Here, the student model obtains the highest accuracy of 0.89 at $T$ values of 74 and 86. In contrast, performance degrades severely at a $T$ of 34, achieving an accuracy of only 0.33. Table \ref{tab:effB3_stud} shows the performance of EfficientNetB3, which is another scaled variant of the EfficientNet family. Using EfficientNetB3 as the teacher and $\alpha$ of 0.4, the student model achieves the highest accuracy of 90\% when $T$ is set to 99, which is higher as compared to EfficientNetB0. On the other hand, the lowest accuracy is 70\% at $T$ of 16.

From the analysis of Tables \ref{tab:resnet_std}, \ref{tab:vgg_student}, \ref{tab:effB0_stud}, and \ref{tab:effB3_stud}, Using ResNet50 as the teacher model with $T$ values of 10 or 40 and an $\alpha$ value of 0.3 results in the best performance of the student model. This specific combination of $T$ and $\alpha$ seems to effectively balance the knowledge transfer process from the teacher, helping to boost the student model's accuracy and generalization capability.

\begin{table*}[ht]
\centering 
\caption{ \centering DCSNet Evaluation with EfficientNetB0 as Teacher}
\renewcommand{\arraystretch}{1.5}
\resizebox{\textwidth}{!}{%
\begin{tabular}{@{}c c c c c c c c@{}}
\toprule
\textbf{Student Model} & \textbf{Trainable Parameter} & \textbf{Alpha} & \textbf{Temperature} & \textbf{Accuracy} & \textbf{Precision} & \textbf{Recall} & \textbf{F1-score} \\
\midrule

\multirow{10}{*}{DCSNet} 
& \multirow{10}{*}{668,931} 
& \multirow{10}{*}{0.4} 
& 7  & 0.75 & 0.78 & 0.75 & 0.74 \\ 
&  &  & 15 & 0.86 & 0.88 & 0.86 & 0.86 \\ 
&  &  & 23 & 0.65 & 0.49 & 0.65 & 0.54 \\ 
&  &  & 34 & 0.33 & 0.11 & 0.33 & 0.17 \\ 
&  &  & 48 & 0.83 & 0.84 & 0.83 & 0.83 \\ 
&  &  & 57 & 0.81 & 0.86 & 0.81 & 0.80 \\ 
&  &  & 65 & 0.80 & 0.85 & 0.80 & 0.78 \\ 
&  &  & 74 & 0.89 & 0.90 & 0.89 & 0.89 \\ 
&  &  & 86 & 0.89 & 0.90 & 0.89 & 0.89 \\ 
&  &  & 98 & 0.87 & 0.89 & 0.87 & 0.86 \\ 

\bottomrule
\end{tabular}%
}
\label{tab:effB0_stud}
\end{table*}

\begin{table*}[ht]
\centering 
\caption{\centering DCSNet Evaluation with EfficientNetB3 as Teacher}
\renewcommand{\arraystretch}{1.5}
\resizebox{\textwidth}{!}{%
\begin{tabular}{@{}c c c c c c c c@{}}
\toprule
\textbf{Student Model} & \textbf{Trainable Parameter} & \textbf{Alpha} & \textbf{Temperature} & \textbf{Accuracy} & \textbf{Precision} & \textbf{Recall} & \textbf{F1-score} \\
\midrule

\multirow{10}{*}{DCSNet} 
& \multirow{10}{*}{668,931} 
& \multirow{10}{*}{0.4} 

& 5  & 0.88 & 0.89 & 0.88 & 0.88 \\ 
&  &  & 16 & 0.70 & 0.79 & 0.70 & 0.65 \\ 
&  &  & 31 & 0.89 & 0.89 & 0.89 & 0.89 \\ 
&  &  & 46 & 0.86 & 0.87 & 0.86 & 0.86 \\ 
&  &  & 57 & 0.85 & 0.87 & 0.85 & 0.84 \\ 
&  &  & 69 & 0.88 & 0.88 & 0.88 & 0.88 \\ 
&  &  & 78 & 0.87 & 0.88 & 0.87 & 0.87 \\ 
&  &  & 86 & 0.87 & 0.88 & 0.87 & 0.87 \\ 
&  &  & 93 & 0.80 & 0.86 & 0.80 & 0.79 \\ 
&  &  & 99 & 0.90 & 0.90 & 0.90 & 0.90 \\ 

\bottomrule
\end{tabular}%
}
\label{tab:effB3_stud}
\end{table*}

Table \ref{tab:resnet_all_student} highlights how DCSNet performs compared to other larger pre-trained models when these models are considered as student networks. Among these pre-trained models, VGG16 achieves the highest accuracy of 0.97, followed by VGG19, ResNet101, and ResNet50, each of 0.96, 0.94, and 0.94, respectively. In contrast, EfficientNetB0 and MobileNetV2 show the lowest accuracy, each achieving 0.67 accuracy. DCSNet achieves an accuracy of 0.92 with significantly fewer parameters, approximately 0.66M, compared to VGG16, which has around 21M parameters, VGG19 with 26M parameters, ResNet101 with approximately 68M parameters, and ResNet50 with 49M parameters. In the case of MobileNetV2, EfficientNetB0, EfficientNetB3, and InceptionV3, which have approximately 18M, 20M, 30M, and 34M parameters, respectively, the proposed model DCSNet, despite having a relatively low parameter count, outperforms these larger models in performance. While several pre-trained students achieve slightly higher accuracy than our DCSNet, the increase in accuracy is small compared to the benefits of efficiency and significantly reduced parameter count that our model offers.

\begin{table*}[ht]
\centering 
\caption{ \centering Teacher Model Evaluation: ResNet50}
\renewcommand{\arraystretch}{1.5}
\resizebox{\textwidth}{!}{%
\begin{tabular}{@{}c c c c c c c@{}}
\toprule
\textbf{Teacher Model} & \textbf{Student Model} & \textbf{Parameter Number} & \textbf{Accuracy} & \textbf{Precision} & \textbf{Recall} & \textbf{F1-score} \\
\midrule

\multirow{9}{*}{ResNet50} 

& EfficientNetB0    & 20,139,430  & 0.67 & 0.80 & 0.67 & 0.57 \\ 
& EfficientNetB3    & 30,084,658  & 0.78 & 0.85 & 0.78 & 0.75 \\ 
& ResNet101         & 68,381,827  & 0.94 & 0.94 & 0.94 & 0.94 \\ 
& VGG16             & 21,170,755  & 0.97 & 0.97 & 0.97 & 0.97 \\ 
& VGG19             & 26,480,451  & 0.96 & 0.96 & 0.96 & 0.96 \\ 
& MobileNetV2      & 18,347,843  & 0.67 & 0.80 & 0.67 & 0.56 \\ 
& InceptionV3      & 34,943,523  & 0.81 & 0.80 & 0.81 & 0.80 \\ 
& ResNet50          & 49,311,363  & 0.94 & 0.95 & 0.94 & 0.94 \\ 
& DCSNet            & 668,931     & 0.92 & 0.92 & 0.92 & 0.92 \\

\bottomrule
\end{tabular}%
}
\label{tab:resnet_all_student}
\end{table*}

\begin{figure}[H]
    \centering
    \includegraphics[width=.42\textwidth]{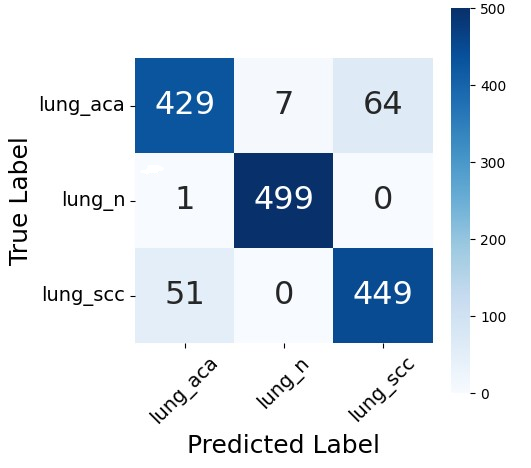}
    \caption{ \centering Confusion matrix of DCSNet model}
    \label{fig: Custom Student Model CM}
\end{figure}

Fig. \ref{fig: Custom Student Model CM} shows the performance of DCSNet in classifying ACA, BN, and SCC. For ACA, 71 FNs are observed, with 64 being misclassified as SCC and seven as BN. FNs arise for SCC with 51, which are misclassified as ACA. The results show that the FN rate for SCC is slightly lower than ACA. Interestingly, the proposed model performs best while classifying healthy samples, with only one BN sample being misclassified as SCC. This suggests that although the model occasionally struggles to differentiate between ACA and SCC, it shows strong robustness in differentiating between malignant and benign tissue.

\section{Explainable AI (XAI)}
\label{sec:XAI}
A significant challenge with machine learning models is their lack of explainability. Transparency is a crucial aspect when it comes to sensitive fields such as healthcare, where the decision-making process must be clear and interpretable. A potential solution to alleviate the black-box nature of models is to use XAI techniques.  

For image data, there are several XAI techniques that help to interpret the model decision-making process. Saliency maps is one such approach; it highlights the pixels or regions most influential in the output of the model. Another method is Class Activation Maps (CAM), such as Grad-CAM, which identify the areas most relevant to the model's decision-making process \cite{selvaraju2017grad}. This approach helps to visualize how certain features contribute towards the model's output. Layer-wise Relevance Propagation (LRP) is another technique that traces the model output back through each layer to find out how each pixel contributes to the decision-making process. Among these techniques, Grad-CAM is particularly useful for understanding CNNs in image-based applications. Grad-CAM generates a heatmap over the input images, highlighting the features most relevant to the model’s decision-making process. This method operates by calculating the gradients of the target output with respect to the activations in the final convolutional layer.

\begin{figure*}[ht]
    \centering
    \includegraphics[width=0.92\textwidth]{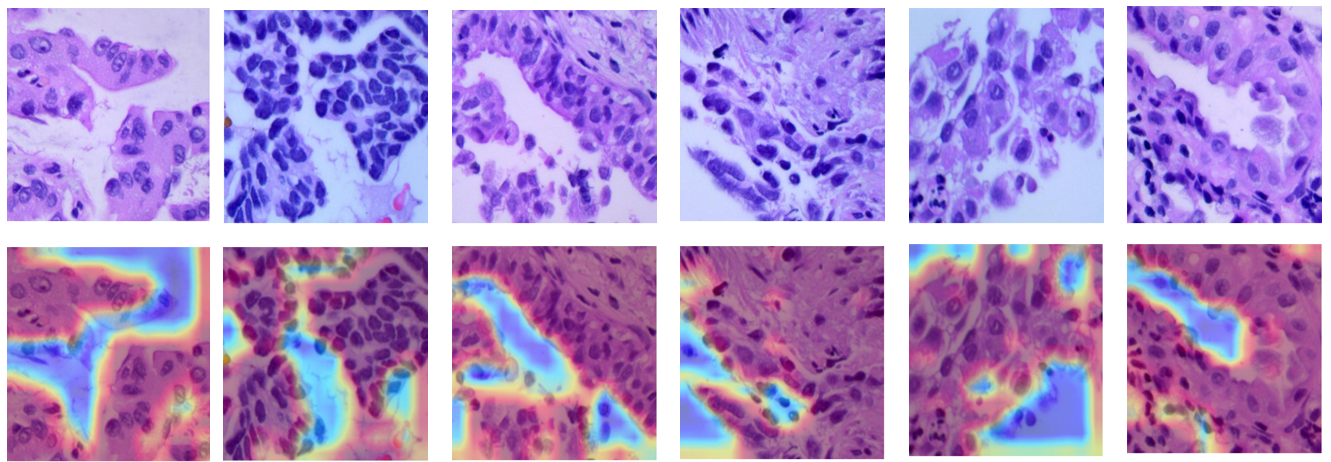} % Use full text width
    \caption{ \centering Grad-CAM visualizations highlighting regions relevant for lung cancer detection}
    \label{fig:gradcam_adc}
\end{figure*}

Fig. \ref{fig:gradcam_adc} illustrates the heatmaps generated by the DCSNet for ACA using Grad-CAM. 
The heatmaps highlight regions where the CNN detects relevant patterns indicating the presence of ACA. The red/orange regions represent areas of high importance, while the blue regions indicate low importance. In Fig. \ref{fig:gradcam_adc}, the heatmaps highlight areas with clusters of cells that may contain abnormal nuclei or glandular patterns. These highlighted areas correspond to common ACA markers. Moreover, the heatmaps avoid irrelevant regions, such as background areas or areas lacking any cellular structures. Based on the Grad-CAM results, we can conclude that the proposed DCSNet model emphasizes only on the regions relevant for the diagnosis of ACA, showcasing its strong performance in identifying key features from histopathological images.\\
The results show that the proposed model can identify ACA accurately and provide insights into the specific regions that drive its decision-making process. Such transparency helps medical experts to validate the model's output.
\section{Conclusion }
\label{sec:conclusion}

This study presents a response-based KD approach for lung cancer detection from histopathological images. The teacher model, ResNet50, is selected based on its superior performance across a variety of pre-trained architectures, which also include ResNet101, EfficientNetB0, EfficientNetB3, VGG16, VGG19, MobileNetV2, and InceptionV3. DCSNet, with a compact architecture consisting of 0.66M parameters, is used as the student model, and the distillation process is carried out under various $T$ scaling and distillation coefficient $\alpha$ settings. The best results are obtained with a $T$ of 10 or 40 and an $\alpha$ value of 0.3, resulting in an impressive accuracy of 0.92 when classifying histopathological images into three distinct categories.

Our approach highlights the efficacy of KD in transferring high-level feature representations from deeper layers of the teacher model to a shallower student model, thereby reducing computational complexity without sacrificing predictive accuracy. In terms of depth, the teacher model ResNet50 is a deep convolutional neural network with 50 layers, leveraging residual blocks that allow the network to learn deeper hierarchical features without suffering from the degradation problem that typically hampers very deep networks. These residual blocks facilitate the direct propagation of gradients through skip connections, enabling the learning of more abstract and high-level features, which are particularly important for complex image recognition tasks like cancer detection in histopathological images.

While ResNet50 provides the depth required to learn intricate spatial patterns and subtle distinctions in the data, DCSNet employs a shallower architecture with fewer parameters to achieve efficient performance. The shallowness of the student network allows it to generalize better to new data by avoiding overfitting, a common challenge in deep neural networks. Furthermore, the distillation process helps DCSNet overcome its initial limitations by transferring the rich feature representations learned by the teacher model in the deeper layers to the student, which compensates for its smaller capacity and promotes more effective learning.

Additionally, comparisons with other student models show that DCSNet surpasses more complex architectures, such as MobileNetV2, EfficientNetB0, EfficientNetB3, and InceptionV3, in terms of performance. To enhance interpretability, Grad-CAM is employed to generate heatmaps, highlighting the regions that influence the model's predictions. The model effectively identifies areas with a high likelihood of carcinoma markers, contributing to the transparency of its decision-making process.

Future work includes exploring advanced KD techniques that leverage both intermediate and final layer knowledge, as well as incorporating model compression strategies like quantization and pruning alongside distillation to develop more efficient and scalable models for lung cancer detection.


\begin{thebibliography}{99}

\bibitem{sutherland2010cell}
Sutherland, K. D., \& Berns, A.: Cell of origin of lung cancer. \textit{Molecular Oncology} \textbf{4}(5), 397--403 (2010). \url{https://doi.org/10.1016/j.molonc.2010.05.002}

\bibitem{a}
American Cancer Society: American Cancer Society Releases Latest Global Cancer Statistics; Cancer Cases Expected to Rise to 35 Million Worldwide by 2050. (April 2024). \url{https://pressroom.cancer.org/GlobalCancerStatistics2024}, Accessed 01 February 2025.


\bibitem{hosseini2024deep}
Hosseini, S. H., Monsefi, R., \& Shadroo, S.: Deep learning applications for lung cancer diagnosis: a systematic review. \textit{Multimedia Tools and Applications} \textbf{83}(5), 14305--14335 (2024). \url{https://doi.org/10.1007/s11042-023-16046-w}

\bibitem{forte2022deep}
Forte, G. C., Altmayer, S., Silva, R. F., Stefani, M. T., Libermann, L. L., Cavion, C. C., Youssef, A., Forghani, R., King, J., Mohamed, T.-L., et al.: Deep learning algorithms for diagnosis of lung cancer: a systematic review and meta-analysis. \textit{Cancers} \textbf{14}(16), 3856 (2022). \url{https://doi.org/10.3390/cancers14163856}

\bibitem{gou2021knowledge}
Gou, J., Yu, B., Maybank, S. J., \& Tao, D.: Knowledge distillation: A survey. \textit{International Journal of Computer Vision} \textbf{129}(6), 1789--1819 (2021). \url{https://doi.org/10.1007/s11263-021-01453-z}

\bibitem{zhang2020improve}
Zhang, L., \& Ma, K.: Improve object detection with feature-based knowledge distillation: Towards accurate and efficient detectors. In: \textit{International Conference on Learning Representations} (2020).

\bibitem{ji2021show}
Ji, M., Heo, B., \& Park, S.: Show, attend and distill: Knowledge distillation via attention-based feature matching. In: \textit{Proceedings of the AAAI Conference on Artificial Intelligence} \textbf{35}(9), 7945--7952 (2021). \url{https://doi.org/10.1609/aaai.v35i9.16969}

\bibitem{park2019relational}
Park, W., Kim, D., Lu, Y., \& Cho, M.: Relational knowledge distillation. In: \textit{Proceedings of the IEEE/CVF Conference on Computer Vision and Pattern Recognition}, pp. 3967--3976 (2019). \url{https://doi.org/10.48550/arXiv.1904.05068}

\bibitem{borkowski2019lung}
Borkowski, A. A., Bui, M. M., Thomas, L. B., Wilson, C. P., DeLand, L. A., \& Mastorides, S. M.: Lung and colon cancer histopathological image dataset (lc25000). \textit{arXiv preprint arXiv:1912.12142} (2019). \url{https://doi.org/10.48550/arXiv.1912.12142}

\bibitem{b1}
Nasser, I. M., \& Abu-Naser, S. S.: Lung cancer detection using artificial neural network. \textit{International Journal of Engineering and Information Systems (IJEAIS)} \textbf{3}(3), 17--23 (2019).

\bibitem{b2}
Al-Tarawneh, M. S.: Lung cancer detection using image processing techniques. \textit{Leonardo Electronic Journal of Practices and Technologies} \textbf{11}(21), 147--158 (2012).





\bibitem{b9}
Shakeel, P. M., Burhanuddin, M. A., \& Desa, M. I.: Lung cancer detection from CT image using improved profuse clustering and deep learning instantaneously trained neural networks. \textit{Measurement} \textbf{145}, 702--712 (2019). \url{https://doi.org/10.1016/j.measurement.2019.05.027}

\bibitem{b10}
Sori, W. J., Feng, J., Godana, A. W., Liu, S., \& Gelmecha, D. J.: DFD-Net: lung cancer detection from denoised CT scan image using deep learning. \textit{Frontiers of Computer Science} \textbf{15}, 1--13 (2021). \url{https://doi.org/10.1007/s11704-020-9050-z}

\bibitem{b7}
Shakeel, P. M., Burhanuddin, M. A., \& Desa, M. I.: Automatic lung cancer detection from CT image using improved deep neural network and ensemble classifier. \textit{Neural Computing and Applications}, 1--14 (2022). \url{https://doi.org/10.1007/s00521-020-04842-6}

\bibitem{b6}
Bhatia, S., Sinha, Y., \& Goel, L.: Lung cancer detection: a deep learning approach. In: \textit{Soft Computing for Problem Solving: SocProS 2017, Volume 2}, pp. 699--705 (2019). \url{https://doi.org/10.1007/978-981-13-1595-4_55}

\bibitem{b4}
Asuntha, A., \& Srinivasan, A.: Deep learning for lung cancer detection and classification. \textit{Multimedia Tools and Applications} \textbf{79}(11), 7731--7762 (2020). \url{https://doi.org/10.1007/s11042-019-08394-3}




\bibitem{b8}
Alakwaa, W., Nassef, M., \& Badr, A.: Lung cancer detection and classification with 3D convolutional neural network (3D-CNN). \textit{International Journal of Advanced Computer Science and Applications} \textbf{8}(8) (2017).

\bibitem{hinton2015distilling}
Hinton, G.: Distilling the knowledge in a neural network. \textit{arXiv preprint arXiv:1503.02531} (2015). \url{https://doi.org/10.48550/arXiv.1503.02531}


\bibitem{b11}
Fukuda, T., Suzuki, M., Kurata, G., Thomas, S., Cui, J., \& Ramabhadran, B.: Efficient knowledge distillation from an ensemble of teachers. In: \textit{Interspeech}, pp. 3697--3701 (2017).


\bibitem{b13}
Yoon, D., Park, J., \& Cho, D.: Lightweight deep CNN for natural image matting via similarity-preserving knowledge distillation. \textit{IEEE Signal Processing Letters} \textbf{27}, 2139--2143 (2020). \url{https://doi.org/10.1109/LSP.2020.3039952}

\bibitem{b12}
Mirzadeh, S. I., Farajtabar, M., Li, A., Levine, N., Matsukawa, A., \& Ghasemzadeh, H.: Improved knowledge distillation via teacher assistant. In: \textit{Proceedings of the AAAI Conference on Artificial Intelligence} \textbf{34}(04), 5191--5198 (2020).

\bibitem{saleknia2024multi}
Saleknia, A. H., \& Ayatollahi, A.: Multi-step knowledge distillation framework for action recognition in still images. In: \textit{2024 20th CSI International Symposium on Artificial Intelligence and Signal Processing (AISP)}, pp. 1--7 (2024). \url{https://doi.org/10.1109/AISP61396.2024.10475221}

\bibitem{romero2014fitnets}
Romero, A., Ballas, N., Kahou, S. E., Chassang, A., Gatta, C., \& Bengio, Y.: Fitnets: Hints for thin deep nets. \textit{arXiv preprint arXiv:1412.6550} (2014). \url{https://doi.org/10.48550/arXiv.1412.6550}


\bibitem{zagoruyko2016paying}
Zagoruyko, S., \& Komodakis, N.: Paying more attention to attention: Improving the performance of convolutional neural networks via attention transfer. \textit{arXiv preprint arXiv:1612.03928} (2016). \url{https://doi.org/10.48550/arXiv.1612.03928}

\bibitem{yang2024vitkd}
Yang, Z., Li, Z., Zeng, A., Li, Z., Yuan, C., \& Li, Y.: ViTKD: Feature-based knowledge distillation for Vision Transformers. In: \textit{Proceedings of the IEEE/CVF Conference on Computer Vision and Pattern Recognition}, pp. 1379--1388 (2024).

\bibitem{b14}
Guo, Z., Yan, H., Li, H., \& Lin, X.: Class attention transfer based knowledge distillation. In: \textit{Proceedings of the IEEE/CVF Conference on Computer Vision and Pattern Recognition}, pp. 11868--11877 (2023).

\bibitem{zhang2019your}
Zhang, L., Song, J., Gao, A., Chen, J., Bao, C., \& Ma, K.: Be your own teacher: Improve the performance of convolutional neural networks via self distillation. In: \textit{Proceedings of the IEEE/CVF International Conference on Computer Vision}, pp. 3713--3722 (2019).


\bibitem{xu2019data}
Xu, T.-B., \& Liu, C.-L.: Data-distortion guided self-distillation for deep neural networks. In: \textit{Proceedings of the AAAI Conference on Artificial Intelligence} \textbf{33}(01), 5565--5572 (2019). \url{https://doi.org/10.1609/aaai.v33i01.33015565}

\bibitem{tu2022general}
Tu, Z., Liu, X., \& Xiao, X.: A general dynamic knowledge distillation method for visual analytics. \textit{IEEE Transactions on Image Processing} \textbf{31}, 6517--6531 (2022). \url{https://doi.org/10.1109/TIP.2022.3212905}


\bibitem{yang2023skill}
Yang, S., Xu, L., Zhou, M., Yang, X., Yang, J., \& Huang, Z.: Skill-transferring knowledge distillation method. \textit{IEEE Transactions on Circuits and Systems for Video Technology} \textbf{33}(11), 6487--6502 (2023). \url{https://doi.org/10.1109/TCSVT.2023.3271124}

\bibitem{zhang2023adaptive}
Zhang, H., Chen, D., \& Wang, C.: Adaptive multi-teacher knowledge distillation with meta-learning. In: \textit{2023 IEEE International Conference on Multimedia and Expo (ICME)}, pp. 1943--1948 (2023). \url{https://doi.org/10.1109/ICME55011.2023.00333}


\bibitem{wang2022nffkd}
Wang, Z., Xie, J., Yao, Z., Kuang, X., Gao, Q., \& Tong, T.: NFFKD: A knowledge distillation method based on normalized feature fusion model. In: \textit{2022 IEEE 5th International Conference on Big Data and Artificial Intelligence (BDAI)}, pp. 111--116 (2022). \url{https://doi.org/10.1109/BDAI56143.2022.9862657}


\bibitem{rao2023parameter}
Rao, J., Meng, X., Ding, L., Qi, S., Liu, X., Zhang, M., \& Tao, D.: Parameter-efficient and student-friendly knowledge distillation. \textit{IEEE Transactions on Multimedia} (2023). \url{https://doi.org/10.1109/TMM.2023.3321480}

\bibitem{qin2021efficient}
Qin, D., Bu, J.-J., Liu, Z., Shen, X., Zhou, S., Gu, J.-J., Wang, Z.-H., Wu, L., \& Dai, H.-F.: Efficient medical image segmentation based on knowledge distillation. \textit{IEEE Transactions on Medical Imaging} \textbf{40}(12), 3820--3831 (2021). \url{https://doi.org/10.1109/TMI.2021.3098703}












\bibitem{zhao2023mskd}
Zhao, L., Qian, X., Guo, Y., Song, J., Hou, J., \& Gong, J.: MSKD: Structured knowledge distillation for efficient medical image segmentation. \textit{Computers in Biology and Medicine} \textbf{164}, 107284 (2023). \url{https://doi.org/10.1016/j.compbiomed.2023.107284}


\bibitem{xing2021categorical}
Xing, X., Hou, Y., Li, H., Yuan, Y., Li, H., \& Meng, M. Q.-H.: Categorical relation-preserving contrastive knowledge distillation for medical image classification. In: \textit{Medical Image Computing and Computer Assisted Intervention--MICCAI 2021: 24th International Conference, Strasbourg, France, September 27--October 1, 2021, Proceedings, Part V}, pp. 163--173 (2021). \url{https://doi.org/10.1007/978-3-030-87240-3_16}


\bibitem{hassan2022knowledge}
Hassan, T., Shafay, M., Hassan, B., Akram, M. U., ElBaz, A., \& Werghi, N.: Knowledge distillation driven instance segmentation for grading prostate cancer. \textit{Computers in Biology and Medicine} \textbf{150}, 106124 (2022). \url{https://doi.org/10.1016/j.compbiomed.2022.106124}



\bibitem{b25}
Wang, S., \& Luo, X.: Incremental learning method for lung nodule detection based on EWC and feature distillation. In: \textit{International Conference on Biomedical and Intelligent Systems (IC-BIS 2022)} \textbf{12458}, 868--874 (2022). \url{https://doi.org/10.1117/12.2660291}



\bibitem{b26}
Zheng, Z., Yao, H., Lin, C., Huang, K., Chen, L., Shao, Z., Zhou, H., \& Zhao, G.: KD\_ConvNeXt: Knowledge distillation-based image classification of lung tumor surgical specimen sections. \textit{Frontiers in Genetics} \textbf{14}, 1254435 (2023). \url{https://doi.org/10.3389/fgene.2023.1254435}


\bibitem{bloice2019biomedical}
Bloice, M. D., Roth, P. M., \& Holzinger, A.: Biomedical image augmentation using Augmentor. \textit{Bioinformatics} \textbf{35}(21), 4522--4524 (2019).


\bibitem{DBLP:journals/corr/HeZRS15}
He, K., Zhang, X., Ren, S., \& Sun, J.: Deep residual learning for image recognition. \textit{CoRR} (2015). \url{https://doi.org/10.48550/arXiv.1512.03385}


\bibitem{DBLP:journals/corr/abs-1905-11946}
Tan, M., \& Le, Q. V.: EfficientNet: Rethinking model scaling for convolutional neural networks. \textit{CoRR} \textbf{abs/1905.11946} (2019). \url{https://doi.org/10.48550/arXiv.1905.11946}


\bibitem{simonyan2014very}
Simonyan, K.: Very deep convolutional networks for large-scale image recognition. \textit{arXiv preprint arXiv:1409.1556} (2014). \url{https://doi.org/10.48550/arXiv.1409.1556}


\bibitem{DBLP:journals/corr/HowardZCKWWAA17}
Howard, A. G., Zhu, M., Chen, B., Kalenichenko, D., Wang, W., Weyand, T., Andreetto, M., \& Adam, H.: MobileNets: Efficient convolutional neural networks for mobile vision applications. \textit{CoRR} \textbf{abs/1704.04861} (2017). \url{https://doi.org/10.48550/arXiv.1704.04861}



\bibitem{DBLP:journals/corr/SzegedyLJSRAEVR14}
Szegedy, C., Liu, W., Jia, Y., Sermanet, P., Reed, S. E., Anguelov, D., Erhan, D., Vanhoucke, V., \& Rabinovich, A.: Going deeper with convolutions. \textit{CoRR} \textbf{abs/1409.4842} (2014). \url{https://doi.org/10.48550/arXiv.1409.4842}


\bibitem{selvaraju2017grad}
Selvaraju, R. R., Cogswell, M., Das, A., Vedantam, R., Parikh, D., \& Batra, D.: Grad-CAM: Visual explanations from deep networks via gradient-based localization. In: \textit{Proceedings of the IEEE International Conference on Computer Vision}, pp. 618--626 (2017).




\end{thebibliography}
\end{document}